\documentclass[twocolumn,amsmath,amssymb,prc,superscriptaddress,floatfix,showpacs,nofootinbib]{revtex4}
\usepackage[dvips]{epsfig}
\usepackage{slashed}
\usepackage{mathrsfs}
\usepackage{graphicx}
\usepackage{xcolor}
\usepackage{hyperref}
\usepackage{dcolumn}
\usepackage{subfigure}
\usepackage{xspace}
\usepackage{bbm}

\definecolor{heidelbeer}{rgb}{0.5,0,0.5}

\graphicspath{
{./}
{./Figures/}
}
\begin{document}
\title{Real-time dynamics of string breaking}

\author{F. Hebenstreit}
\affiliation{Institut f\"{u}r Theoretische Physik, Universit\"{a}t Heidelberg,
  Philosophenweg 16, 69120 Heidelberg, Germany}

\author{J. Berges}
\affiliation{Institut f\"{u}r Theoretische Physik, Universit\"{a}t Heidelberg,
  Philosophenweg 16, 69120 Heidelberg, Germany}
\affiliation{ExtreMe Matter Institute EMMI, GSI Helmholtzzentrum,
  Planckstra\ss e 1, 64291 Darmstadt, Germany}

\author{D. Gelfand}
\affiliation{Institut f\"{u}r Theoretische Physik, Universit\"{a}t Heidelberg,
  Philosophenweg 16, 69120 Heidelberg, Germany}

\begin{abstract}
We study the real-time dynamics of string breaking in quantum electrodynamics in one spatial dimension. 
A two-stage process with a clear separation of time and energy scales for the fermion--antifermion pair creation and subsequent charge separation leading to the screening of external charges is found. 
Going away from the traditional setup of external static charges, we establish the phenomenon of multiple string breaking by considering dynamical charges flying apart.
\end{abstract}
\maketitle


The string formation between an external static quark and an antiquark is an important manifestation of the physics of confinement in the theory of strong interactions (QCD). 
In general, in theories with dynamical fundamental charges the confining string can break because of the creation of charge-anticharge pairs which screen the static sources \cite{Bock,Knechtli,Philipsen:1998de,Gliozzi,Bali:2005fu,Pepe:2009in}. 
In particular, quantum electrodynamics (QED) in one spatial dimension shares the nonperturbative phenomenon of string breaking by dynamical fermion-antifermion pair creation with QCD. 

Our current understanding of string breaking mainly concerns static properties obtained from equilibrium lattice Monte Carlo simulations.
These equilibrium calculations can be based on a Euclidean formulation, where the time variable is analytically continued to imaginary values. 
However, in real time this phenomenon can be a process far from equilibrium with a hierarchy of time scales, which is not amenable to a Euclidean formulation. 
Recently, the prospect of constructing quantum simulators for gauge theories with fermions using ultra-cold atoms in an optical lattice~\cite{Banerjee:2012pg,Zohar:2013zla,TCZM} boosted the interest in the real-time dynamics of string breaking. 
First computations in this context concentrate on quantum link models~\cite{Banerjee:2012pg,Marcos:2013aya} and it is an important task to extend these investigations to QED and QCD. 

In this work we present for the first time a detailed {\it space-time picture of string breaking} in QED in one spatial dimension. 
This is possible since in this case the quantum dynamics of string breaking can be accurately mapped onto a classical problem, which can be rigorously solved on a computer using lattice gauge theory techniques~\cite{Aarts:1998td,Berges:2010zv,Hebenstreit:2013qxa}. 
For the case of two external static charges we establish a {\it two-stage process}: Exceeding a critical distance between the external charges quickly leads to spontaneous creation of fermion--antifermion pairs. 
However, the dynamical charges are produced on top of each other and, therefore, initially do not screen the external charges. 
We find that it takes a much longer time to separate the dynamical charges such that the string can finally break. 
Strikingly, it turns out that most of the energy content of the string goes into the work that is required for the process of charge separation, and only a small fraction is spent on pair creation. 
This has a significant impact on the estimate of the critical charge separation for string breaking, and we give a simple model that explains our simulation results. 
We then exploit the rich phenomenology that becomes accessible in a real-time treatment of string formation and subsequent breaking. 
For this purpose, we discard external charges and consider the physical situation of dynamical charges only. 
This allows us to establish the phenomenon of multiple string breaking from dynamical charges flying apart.  
 
The vacuum of QED is unstable against the formation of many-body states in the presence of strong electric fields. 
The creation of electron-positron pairs in a uniform electric field may be viewed as a quantum process in which virtual electron-positron dipoles can be separated to become real pairs once they gain the binding energy of twice the rest mass, $2m$, where we use the convention with a speed of light equal to one.
This Schwinger process is exponentially suppressed unless a critical field strength determined by the electron mass $m$ and the electric charge $e$ is reached~\cite{Sauter:1931zz,Heisenberg:1935qt,Schwinger:1951nm}:
\begin{equation}
\label{eq:Ec}
E_{\rm c} = \frac{m^2}{e} \, .
\end{equation}
 
For a confining string connecting two external static charges, the energy content of the string rises linearly with the distance between the charges. 
For the case of QED in one spatial dimension with $N_0$ external charges $\pm eN_0$ that are separated by a distance $d$, Gauss' law \mbox{$\partial_x E=eN_0\left[\delta\!\left(x+d/2\right)-\delta\!\left(x-d/2\right)\right]$} results in a homogeneous electric field $E_{\rm str}=eN_0$ between the two charges.  
Accordingly, the potential energy rises linearly with the separation $d$:
\begin{equation}
\label{eq:energy}
 V_{\rm str}=\frac{1}{2}\int_{-d/2}^{d/2}{dx \, E_{\rm str}^2}=\frac{e^2N_0^2d}{2} \ .
\end{equation}
In the absence of dynamical fermions, this equation holds for arbitrary separations $d$. 
However, in the interacting theory fermion--antifermion pairs will be created spontaneously once the energy content of the string becomes large enough for distances exceeding a critical distance $d_{\bf c}$. 
As a dynamical process, string breaking can be defined to happen at the time when the total screening of the external charges by the dynamically created pairs occurs such that the corresponding electric field vanishes. 
For this it is necessary to produce at least $N_0$ fermion--antifermion pairs.
Due to the exponential suppression of the Schwinger mechanism, this is expected to occur efficiently only for $E_{\rm str} \gtrsim m^2/e$ according to (\ref{eq:Ec}).
Therefore, we consider $e/m=1/\sqrt{N_0}$ in the following such that $E_{\rm str}=E_{\rm c}$. 
Below we will discuss also more general sets of parameters in the context of multiple string breaking. 

We compute this process from first principles using real-time simulation techniques for lattice QED with Wilson fermions following Refs.~\cite{Aarts:1998td,Berges:2010zv,Hebenstreit:2013qxa}. 
In this nonperturbative approach the full quantum dynamics of fermions is included while the gauge field dynamics is accurately represented by classical simulations for relevant field strengths. 
The real-time simulations are performed on a spatial lattice with the number of sites ranging from $1024$ up to $4096$ and lattice spacings between $a_s=0.05/m$ and $0.1/m$, with temporal steps $a_t/a_s = 0.0125$ -- $0.04$.
We carefully checked the insensitivity of our results to volume and lattice spacing variations.
Observables such as the charge density $\rho(x,t)$ or the fermion density $n(x,t)$ are calculated from gauge-invariant correlation functions in a standard way~\cite{Hebenstreit:2013qxa}. 
Here the fermion density $n(x,t)$ is related to the fermion energy density so that fermions and antifermions contribute with the same sign.
As these observables are defined from the quantum expectation value of correlation functions, quantities like the average number of fermion--antifermion pairs $N(t)=\int{dx\, n(x,t)}/2$ can take on non-integer values.

\begin{figure}[t!]
 \centering
 \includegraphics[width=0.97\columnwidth]{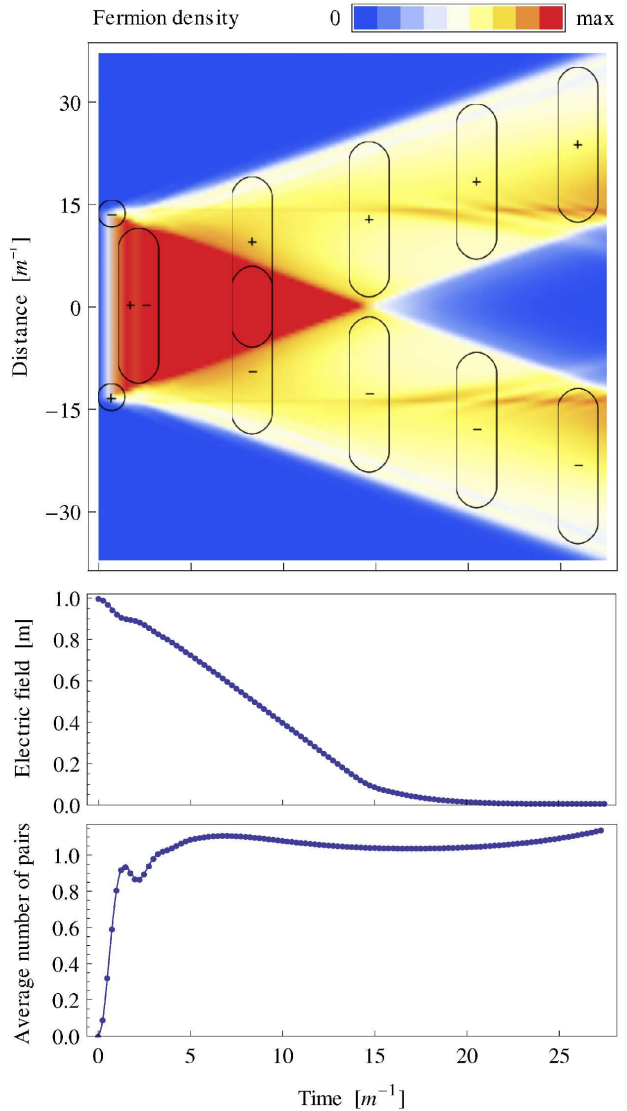}
 \caption{Space-time evolution of string breaking for external static charges $\pm e$ (denoted by $\ominus$ and $\oplus$) with $e/m=1$ separated by $d=28/m$. 
 {\bf Top:} Fermion density $n(x,t)$. The vertical ovals represent the charge density $\rho(x,t)$ according to our model (\ref{eq:charge}) for charge separation. The charge density vanishes in regions where positively and negatively charged ovals overlap.
 {\bf Middle:} Time-dependence of the electric field $E(t)$ at $x=0$. 
 {\bf Bottom:} Average number of fermion--antifermion pairs $N(t)$.}
 \label{fig:static}
\end{figure}

First, we consider the case $N_0=1$ such that $e/m=1$.
In Fig.~\ref{fig:static} the space-time evolution of the fermion density $n(x,t)$ is shown for two external static charges $\pm e$ separated by $d=28/m$, along with the electric field $E(t)$ at $x=0$ as well as the average number of pairs $N(t)$ as a function of time.
From the simulations we find that the employed separation of external charges just lies above the required critical distance $d_{\bf c}$ for string breaking.  
At early times, the fermion density $n(x,t)$ between the external charges increases due to the Schwinger mechanism on rather short time scales of $t_{\rm prod}\simeq 1/m$. 
At the same time, we find that the charge density still vanishes, $\rho(x,t)=0$: 
Fermions and antifermions are initially produced on top of each other and, accordingly, the dynamically created charges do not screen the electric field $E_{\rm str}$ yet. 
After the first stage, fermion--antifermion production has ceased and the average number of pairs $N(t_{\rm prod})\simeq N_0$ stays practically constant. 
At the same time, the remaining electric field separates the dynamically created charges, which is a much longer lasting process with a separation time $t_{\rm sep}\simeq 20/m$. 
Due to the continuous separation process, the external charges are gradually screened so that $E(t)\to 0$ in the end. 
This screening process shows a linear behavior in time since the dynamically created charges move apart from each other close to the forward light cone.
Remarkably, only a rather small fraction of the initial electric field energy is expended on the rest mass energy, $V_{\rm str}>2m$, whereas the largest fraction is used for separating the charges.

We have also simulated the system in the weak coupling regime $e/m=1/\sqrt{N_0}$ with $N_0=2,3,4,5$ such that still $E_{\rm str}=E_{\rm c}$.
The picture of a two-stage process is seen also in these cases with the critical distance showing the dependence $d_{\bf c}\simeq26/e=26\sqrt{N_0}/m$.

We now give a simple dynamical picture providing, in particular, semi-quantitative estimates for $d_{\bf c}$ as well as the charge separation work $W$. 
To describe the fermion--antifermion production, we employ a model which is based on the one-dimensional Schwinger formula, which is typically applicable even for slowly varying electric fields:
\begin{equation}
 \label{eq:schwinger}
 \dot{N}(t)=d\frac{eE(t)}{2\pi}\exp\left(-\frac{\pi m^2}{eE(t)}\right) \ ,
\end{equation}
with $N(0)=0$.
For $t\lesssim t_{\rm prod}\ll d$, the electric field $E(t)$ decreases with time due to the production of fermion--antifermion pairs as well as the gradual screening of the external charges.
In this regime, the field can be approximately described by
\begin{equation}
 \label{eq:efld}
 E(t)\simeq\sqrt{e^2N_0^2-\frac{4mN(t)}{d}}-\frac{eN(t)}{d}t \ .
\end{equation}
Solving the differential equation (\ref{eq:schwinger}) with (\ref{eq:efld}), such that $N(t_{\rm prod})\to N_0$, results in a numerical estimate for the critical distance $d_{\bf c}\simeq28.5/e=28.5\sqrt{N_0}/m$, which is in good agreement with the values we find in our real-time lattice simulations.

Moreover, we give an estimate of the charge separation work $W$ which is based on the simple model that two homogeneous regions of positive and negative charge density are produced on top of each other at some time $t=0$.
These charges are then accelerated by the electric field and move apart from each other close to the forward light cone:
\begin{equation}
 \label{eq:charge}
\rho^{\pm}(x,t)=\pm\frac{eN_0}{d_{\bf c}}\left[\Theta\!\left(x\mp t+\frac{d_{\bf c}}{2}\right)-\Theta\!\left(x\mp t-\frac{d_{\bf c}}{2}\right)\right] ,
\end{equation}
with the average charge $\int{dx \rho^{\pm}(x,t)}=\pm eN_0$ and $\Theta(x) = 1$ for $x > 0$ while being zero otherwise. 
For this model, by applying Gauss' law, the electric field $E(x,t)$ is obtained analytically.
The work done by the electric field on the positive and negative charges upon separating them over a distance $d_{\bf c}/2$, such that the electric field is completely screened at $x=0$, is then given by
\begin{equation}
\label{eq:work}
W^{\pm}=\pm \frac{eN_0}{d_{\bf c}}\int_{-d_{\bf c}/2}^{d_{\bf c}/2}dx_i\int_{x_i}^{x_i\pm d_{\bf c}/2}dxE(x,t)=\frac{5e^2N_0^2 d_{\bf c}}{24} \ ,
\end{equation}
where the integral is over the time-dependent paths $x(t)= x_i + t$.
Plugging $d_{\bf c}$ into the expression for the work (\ref{eq:work}) one obtains: 
\begin{equation}
 W=W^++W^-=\frac{5e^2N_0^2d_{\bf c}}{12}\simeq12mN_0^{3/2} \ .
\end{equation}
This confirms our findings that the total work for charge separation well exceeds the rest mass energy $2mN_0$.

The two-stage process of fermion--antifermion production and charge separation describes the early-time behavior of the system well.
At later times, however, the picture becomes more involved due to the dynamics of the created fermion--antifermion pairs coupled to the electric field.
Here, we want to mention two effects which can be observed at later times: screening of external charges and propagating charge-neutral states.

\begin{figure}[t!]
 \centering
 \includegraphics[width=0.97\columnwidth]{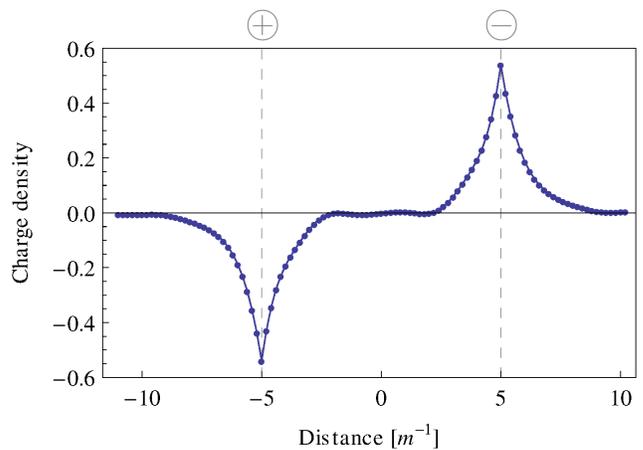}
 \caption{Screening of external charges. The charge density $\rho(x,t=100/m)$ is shown for external static charges $\pm e$ (denoted by $\ominus$ and $\oplus$) separated by $d=10/m$ for $e/m=2$.}
 \label{fig:long_time}
\end{figure}

In Fig.~\ref{fig:long_time} the charge density $\rho(x,t=100/m)$ is shown at late times for $N_0=1$ separated by $d=10/m$ in the strong-coupling regime with $e/m=2$, such that $E_{\rm str}=4E_{\rm c}$.
At early times, we again observe the two-stage process of pair production and charge separation.
However, due to the particular choice of $d$ and $E_{\rm str}$ there are more than one but rather $N(t_{\rm prod})\simeq5$ fermion--antifermion pairs produced. 
Accordingly, only one fermion and antifermion are subsequently used to screen the external charges $\pm e$. 
For this configuration we find for its spread $\simeq3/m$.
This behavior resembles the screening of external charges in the Schwinger model, corresponding to the limit $e/m\to\infty$ \cite{Iso:1988zi}.
The remaining $4N(t)/5\simeq4$ fermion--antifermion pairs, however, bunch to composite charge-neutral states which propagate freely since the external charges are totally screened.
A detailed description of this effect is deferred to a future investigation.

\begin{figure}[t!]
 \centering
 \includegraphics[width=0.97\columnwidth]{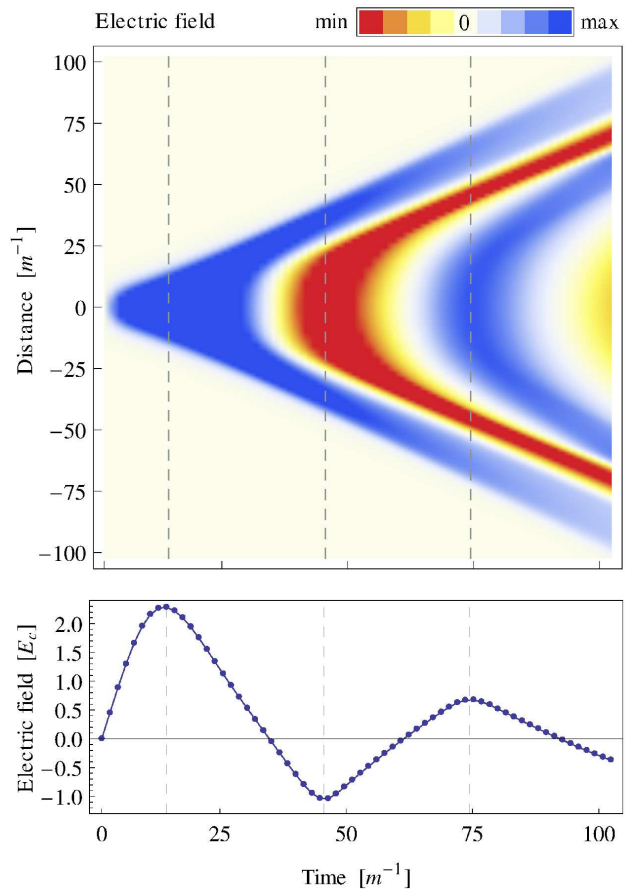}
 \caption{Space-time evolution of multiple string breaking from dynamical charges flying apart. {\bf Top:} Electric field $E(x,t)$. {\bf Bottom:} Central electric field $E(0,t)$ in units of $E_{\rm c}$. The dashed lines indicate the times at which $E(0,t)$ is extremal.}
 \label{fig:dyn_electric}
\end{figure}

\begin{figure}[t!]
 \centering
 \includegraphics[width=0.97\columnwidth]{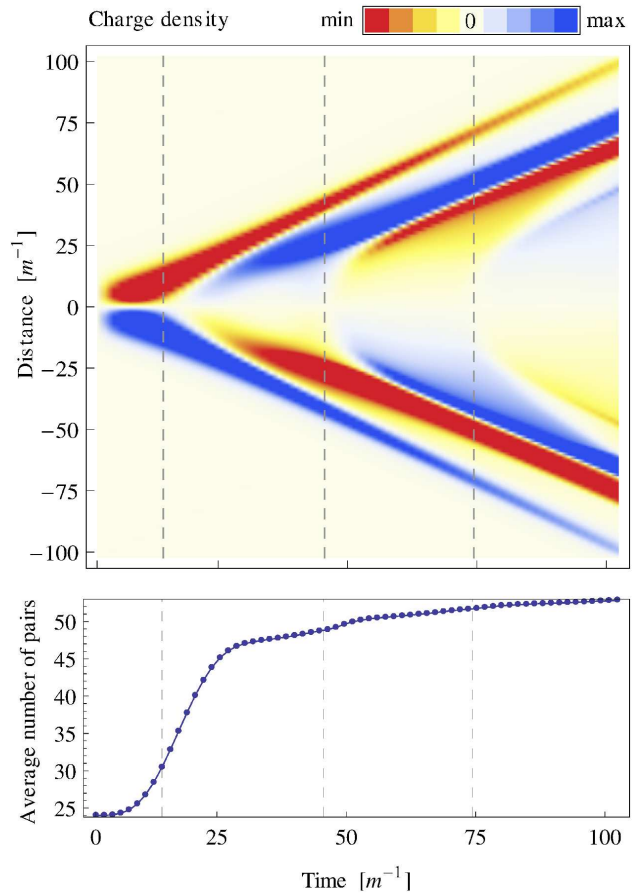}
 \caption{{\bf Top:} Charge density $\rho(x,t)$. {\bf Bottom:} Average number of fermion--antifermion pairs $N(t)$. The dashed lines indicate as in Fig.~\ref{fig:dyn_electric} the times at which $E(0,t)$ becomes extremal.}
 \label{fig:dyn_charge}
\end{figure}

So far we considered string breaking for two external static charges.
We now generalize the above setup by simulating two oppositely charged bunches of dynamical fermions moving apart from each other, i.~e.\ we no longer include external static charges. 
These bunches can be either produced by an external field pulse, or, more directly, one can initialize the fermion fields according to a given distribution \cite{Hebenstreit:2011wk}. 
Here we employ Gaussian distributions around $x=0$ with a width of $\sigma_x = 5/m$ in real-space and $\sigma_p = 4.6 m$ in momentum-space. 
We initialize two fermion bunches with relativistic momenta in opposite direction with an initial number of pairs $N(0)=24$ and given coupling $e/m=0.35$. 

In order to visualize the time evolution, we display in Fig.~\ref{fig:dyn_electric} the electric field $E(x,t)$ (upper panel) and its value at $x=0$ (lower panel). 
Moreover, we show in Fig.~\ref{fig:dyn_charge} the charge density $\rho(x,t)$ (upper panel) and the average number of pairs $N(t)$ (lower panel).
Due to the initial relativistic momenta of the fermions and antifermions, they move apart from each other with a velocity close to the speed of light. In the current configuration, fermions with negative/positive charge move into the positive/negative $x$--direction. 
Upon separating from each other, an electric field string is formed between them. 
For the chosen initial conditions the maximum achieved field strength is much larger than $E_{\rm c}$. 
The time at which this maximum is reached is indicated by the first dashed line in Figs.~\ref{fig:dyn_electric} and \ref{fig:dyn_charge}. 
Around this time, efficient fermion production sets in such that the average number of pairs $N(t)$ rises significantly. 
In complete analogy to the above discussion, the newly created charges still sit on top of each other such that the electric field is not screened yet.

In order to screen the initial bunches, the newly created charges need to be separated.
As a consequence, the electric field performs work and drops linearly with time and finally even changes sign.
At that time, two new bunches of fermions have formed which are oppositely charged compared to the initial ones, and again move apart from each other ({\it primary string breaking}). 
This results in a secondary electric string with a maximum field strength of the order of $-E_{\rm c}$, indicated by the second dashed line. 
As a consequence, fermion production sets in again, however, less efficient than before because of the lower maximum field strength. 
Charges are again created on top of each other and are subsequently separated, resulting in a rise of the electric field including a sign change. 
As a consequence, the formation of two new fermion bunches can be observed, again oppositely charged compared to the previous ones ({\it secondary string breaking}). 
The following extremum of the electric field, as indicated by the third dashed line in the corresponding figures, is already below the critical field strength such that fermion production effectively stops and the average number of pairs becomes asymptotically constant.


To conclude, our results provide unprecedented insights into the real-time dynamics of string formation and breaking from first principles. The described phenomenon of string breaking in QED is intimately related to a one-dimensional geometry, which poses strong constraints on possible experimental realizations. 
However, ultracold atoms in an optical lattice could provide a perfect laboratory for this type of physics, in particular, since they are very suitable to access low-dimensional geometries. 
For the specific case of QED in one spatial dimension, one can use angular momentum conserving atomic scattering processes to directly implement the $U(1)$ gauge symmetry without the need to construct low-energy effective theories~\cite{Zohar:2013zla}. 
In this context, our calculation serves as an important validator for quantum simulators using cold atoms. 
  

\textbf{Acknowledgements:} We thank B.-J.~Sch\"afer for inspiring discussions. 
D.~Gelfand thanks HGS-HIRe for FAIR for support. 
F.~Hebenstreit is supported by the Alexander von Humboldt Foundation.


\end{document}